# Epsilon-near-zero regime as the key to ultrafast control of functional properties of solids

M. Kwaaitaal[1,2,*], D.G Lourens[1,2], C.S. Davies[1,2], and A. Kirilyuk[1,2,*]

[1]*FELIX Laboratory, Radboud University, Toernooiveld 7, 6525 ED Nijmegen, The Netherlands*
[2]*Radboud University, Institute for Molecules and Materials, Heyendaalseweg 135, 6525 AJ Nijmegen, The Netherlands*

*Corresponding authors. Email: marten.kwaaitaal@ru.nl and andrei.kirilyuk@ru.nl

**Strong light-matter interaction constitutes the bedrock of all photonic applications, empowering material elements with the ability to create and mediate interactions of light with light. Amidst the quest to identify new agents facilitating such efficient light-matter interactions, a class of promising materials have emerged featuring highly unusual properties deriving from their dielectric constant $\varepsilon$ being equal, or at least very close, to zero[1,2,3]. Works so far have shown that the enhanced nonlinear optical effects displayed in this 'epsilon-near-zero' (ENZ) regime makes it possible to create ultrafast albeit transient optical switches[4,5]. An outstanding question, however, relates to whether one could use the amplification of light-matter interactions at the ENZ conditions to achieve *permanent* switching. Here, we demonstrate that an ultrafast excitation under ENZ conditions can induce permanent all-optical reversal of ferroelectric polarization between different stable states. Our reliance on ENZ conditions that naturally emerge from the solid's ionic lattice, rather than specific material properties, suggests that the demonstrated mechanism of reversal is truly universal, being capable of permanently switching order parameters in a wide variety of systems.**

The spontaneous polarization is a critically important order parameter already underpinning a broad range of electronic, optical, and electro-mechanical appliances, with clear potential to produce novel functionalities[6] ranging from high-density non-volatile memories[7,8] and extremely sensitive nano-devices[9] to neuromorphic-inspired computational schemes[10] and energy harvesting[11]. All these applications stem from the capacity of electric fields to switch polarized domains in ferroelectric materials[12]. Moreover, the possibility of coupling the polarization of a ferroelectric with other physical parameters such as strain, magnetic order or optical properties provides a unique basis for creating materials that are truly multifunctional[13]. The development of ultrafast and energy-efficient methods capable of coherently switching spontaneous polarization is therefore vital for the future progress in developing new ferroelectric technologies.

As the model system for our investigation, we select one of the most studied classical ferroelectrics, barium titanate ($BaTiO_3$). Barium titanate is a multiaxial perovskite ferroelectric[14] with a cubic parent phase that becomes tetragonal upon the appearance of the ferroelectric polarization **P**. At the transition, the titanium and oxygen ions experience small opposing shifts along one of the cubic axes[15], lowering $BaTiO_3$'s point group symmetry from $m\bar{3}m$ to $4mm$[14]. Such symmetry breaking creates a ferroelectric polarization along one of the crystal axes. In total this results in six possible directions of **P**, thus supporting the stable existence of both 90° and 180° ferroelectric domains[16]. We employ single-domain crystals of BaTiO3 at room temperature, to study how their domain structures are impacted by an excitation at ENZ conditions.

In the spectral range of 7-28 μm (357-1429 cm$^{-1}$), BaTiO3 has two major phonon resonances. Both show significant splitting of transverse and longitudinal optical modes (TO and LO, respectively) due to the strong polar character of the crystal[17]. The LO-phonon frequency is marked by the strongly revealed ENZ regime, where both the real $\varepsilon_1$ and imaginary $\varepsilon_2$ parts of the dielectric constant approach zero at the same time. Moreover, the LO mode has considerable spectral separation from the corresponding IR-active TO mode, which is instead indicated by a maximum in $\varepsilon_2$. The fact that such 'phononic' ENZ materials require no special efforts to construct, while still leading to the best possible realization of ENZ conditions, arguably offers a strong advantage compared to carefully-designed metamaterials[5,18].

To reach the most efficient interaction of the infrared light with the given material, it is imperative to use ultrashort and intense excitation pulses that are sufficiently narrow in bandwidth to ideally fit the ENZ condition[4]. The only source that is capable of meeting this is a free-electron laser, which is well-known to produce pulses that intrinsically obey the Fourier-transform-limit relation between the pulse duration and bandwidth[19]. In this work, we use radiation with wavelengths in the range of 7-28 μm delivered by the free-electron laser facility FELIX in Nijmegen, The Netherlands. FELIX provides transform-limited pulses that are 1-3 ps-long, with 0.5-1% of bandwidth, and with pulse energies exceeding 10 μJ.

After preparing BaTiO3 in a single-domain state, we expose it to an infrared narrow-band excitation with a central wavelength of $\lambda = 14$ μm. Second-harmonic-generation microscopy, using near-infrared incident light, reveals the light-induced creation of 180° domains that are absent at the center of the laser pulse but rather manifest at the sides across the horizontal axis (see Fig. 1(A)). Polarizing microscopy (Fig. 1(B)) shows that this infrared excitation also generates narrow 90° domains that extend across the center of the irradiated region, as well as a set of circular rings. For clarity, we have shown in Fig. 1 two separate images that show the creation of one specific domain type. In general, however, it is possible to have 90° and 180° domains created at the same time (Supplementary Information). Note that such domains can be created both by a single infrared ps-long pulse or an 8-μs-long burst of pulses coming at a repetition rate of 25 MHz (Supplementary Information). Since the latter "macropulse" typically produces larger switched domains, which are therefore more stable, we generally used macropulses as a pump for convenience.

We first consider the concentric rings made visible by the polarizing microscopy in Fig. 1(B). We find that the rings persist upon tuning the pumping wavelength, while the number of rings increase with the optical fluence. The spectral indifference of this feature allows us to conclude that this originates from incoherent heating of the sample, which causes a thermally-induced variation in birefringence (Supplementary Information). This feature is rather useful in providing an in-situ "thermometer" at the sample surface, while also proving that the switched 90° and 180° domains are completely unassociated with heating effects (Supplementary Information).

So far, the switching of ferroelectric domains was achieved with a pumping wavelength of $\lambda = 14$ μm and rms bandwidth of 70 nm, where the ENZ condition is ideally satisfied (see Fig. 2). To therefore assess the importance of the excitation's wavelength, we adjusted $\lambda$ while continuously monitoring the sample's polarization using both SHG and polarizing microscopy.

At varying wavelengths, both the 90° and 180° domains appeared larger or smaller, allowing us to calculate the area of the switched domain. In Fig. 2(A), we plot the spectral dependence of the switched area (normalized by the incident pulse energy), revealing two maxima in the vicinity of 14 μm and 21 μm. These are exactly the wavelengths where the real part of the dielectric function crosses zero and the imaginary part is close to zero. At these wavelengths, the switching of the sufficiently-large ferroelectric domains was permanent, whereas at other pumping wavelengths the switched domains of smaller size were transient, shrinking and vanishing on the timescale of milliseconds (see Supplementary Material). Notice in the spectrum that the success of switching depends on both $\varepsilon_1$ *and* $\varepsilon_2$ going close to zero at the same time.

The spectral dependence of domain switching in Fig. 2 clearly and unambiguously shows the importance of using an excitation at the ENZ conditions. How can we actually explain this though? In fact, a considerable amount of research has already been devoted to formulating and realizing a mechanism that would provide the means for ultrafast all-optical switching of ferroelectric polarization. The first proposal in this direction was the prediction that a high-amplitude excitation of the ferroelectric soft mode can lead, at certain conditions, to the full reversal of the polarization[20]. Subsequent research has revealed that a nonlinear coupling of different optical phonon modes provides a more efficient mechanism involving the displacement of the equilibrium atomic positions that actually deforms the crystal lattice[21,22]. In this process, a strong excitation of an IR-active TO phonon mode along the coordinate $\Omega_{IR}$ results, via a nonlinear coupling to a Raman-active mode, in the shift of the equilibrium position along the latter's coordinate $\Omega_R$. Such a kind of interaction is described by high-order energy terms of the phonon coordinates such as $\Omega^2_{IR}\Omega_R$ or $\Omega^2_{IR}\Omega^2_R$, appearing in the perturbative part of the Hamiltonian. Non-linear phononics has indeed been able to demonstrate a transient switching of ferroelectric polarization[23], albeit only to a small extent and during an extremely short timescale of just 200 fs.

In our experiment however, at 14 μm, no IR-active phonon is present at this excitation frequency, and the LO-mode that is present cannot be excited with a transverse electromagnetic wave at normal incidence, as used in our experiment. Then, if not a phonon, what do we do excite? Deriving from the ENZ conditions, several factors provide for the strong interaction of the light's electric field with the material. These include diverging wavelengths leading to an in-phase excitation across a large volume as well as a vanishing group velocity that increases the interaction time[24,25]. Consequently, a strong nonlinear polarization is created at the excitation frequency and over a significant volume of the sample. Note that the transverse character of the excitation is not a restrictive factor anymore, because of this nonlinearity[4]. This substantially helps given the fact that at the ENZ condition, only (near-)normal incidence results in light penetrating the medium[26,27]. However, how does this non-resonant oscillatory polarization leads to a structural breaking of symmetry?

In fact, it is the combination of the strong induced polarization with the incident electric field that can also lead to a unidirectional displacement of Raman phonons. Helped by the resonance with an IR-active mode, such an effect was shown to be several orders of magnitude stronger than the nonlinear phononic one discussed above[28]. Unlike in the nonlinear phononics, here the resulting nonlinear ionic polarizations in combination with the incident electric fields drive the main effect, that both are amplified by the ENZ condition.

Regardless of the exact microscopic mechanism and the particular Raman mode excited, the pattern of the created domains can still be phenomenologically explained by taking into account the macroscopic shape of the laser pulse. The IR pulse was focused on the sample at normal incidence, resulting in a Gaussian profile of intensity with a FWHM in excess of 100 μm. The magnitude of the atomic displacements scales directly with the light-delivered electric field, and thus the deformation profile will also be Gaussian. By virtue of ferroelectrics being piezoelectric, the Gaussian deformation will similarly induce an electric displacement field. The direction and shape of the latter matches that of the created ferroelectric domains. This entire process is visualized schematically in Fig. 3(A)-(B), with the resulting calculated piezoelectric displacement field shown in Fig. 3(C). A quantitative description is provided in the Methods.

To summarize, we have identified that an excitation under ENZ conditions enables permanent switching of ferroelectric polarization in different directions. The universality of this effect is underscored by the recent observation of switching of magnetic order in ferrimagnetic iron garnets[29] and antiferromagnetic nickel oxide[30], where in both cases it was the excitation at the LO phonon frequency that led to the domain switching or displacement (though the mechanism was not identified). Thus, very different crystallographic symmetries lead to the same effect, adding to the certainty that it is the nonlinear optics, and not a combination of very specific mode symmetries, that is at play here. Looking ahead, further experimental campaigns must explore the dynamics of the switching process with the aim to identify the timescale of the transient response. Nonlinear polarizations, for example, depend sensitively on the timescale of the pulse, whereas nonlinear phononics involves effects that depend on the lifetime of the excited IR phonon. This temporal fingerprint will provide invaluable insight in to the microscopic mechanisms involved in the ENZ regime.

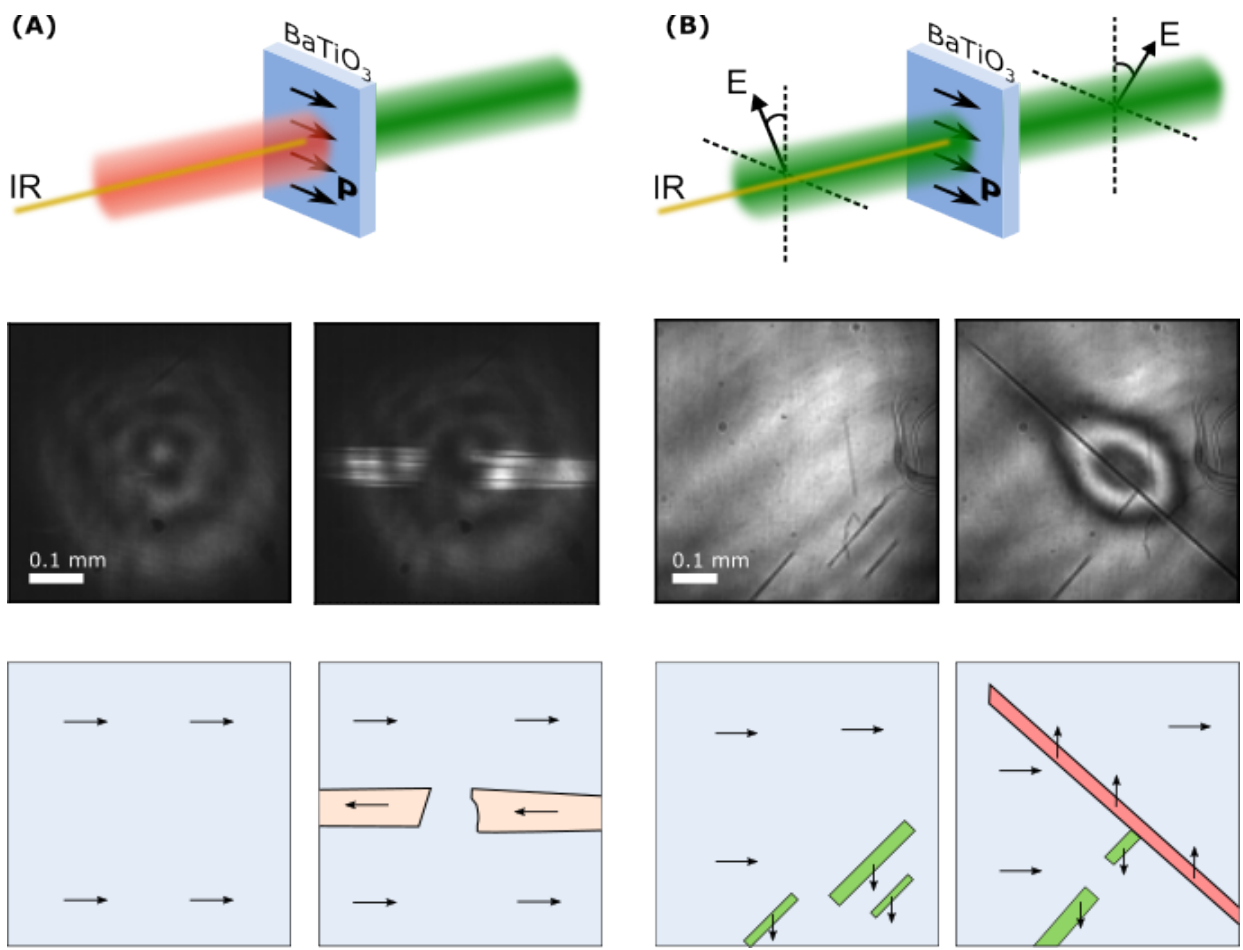

**Fig. 1: Light-induced switching of polarization.**
(**A**)-(**B**) Top panel: Schematic of the experimental setup. A ferroelectric BaTiO₃ sample with polarization **P** is excited by an infrared (IR) pulse at normal incidence (wavelength $\lambda$ = 7-28 μm, gold line). The impact of the IR pulse on the polarization is detected in one of two ways. Column (A): Second-harmonic-generation microscopy resolves 180° switched domains via the polarization-dependent frequency-doubling of near-IR ($\lambda$ = 1040 nm) probe pulses. Column (B): Polarizing microscopy resolves 90° switched domains through the polarization-dependent rotation of the probe's electric field **E**. Middle panel: Typical raw images of the ferroelectric polarization taken before and shortly after illumination of the sample with an IR pump pulse ($\lambda$ = 14 μm). Bottom panel: Schematic distribution of the ferroelectric polarization as deduced from the change of contrast in the recorded images.

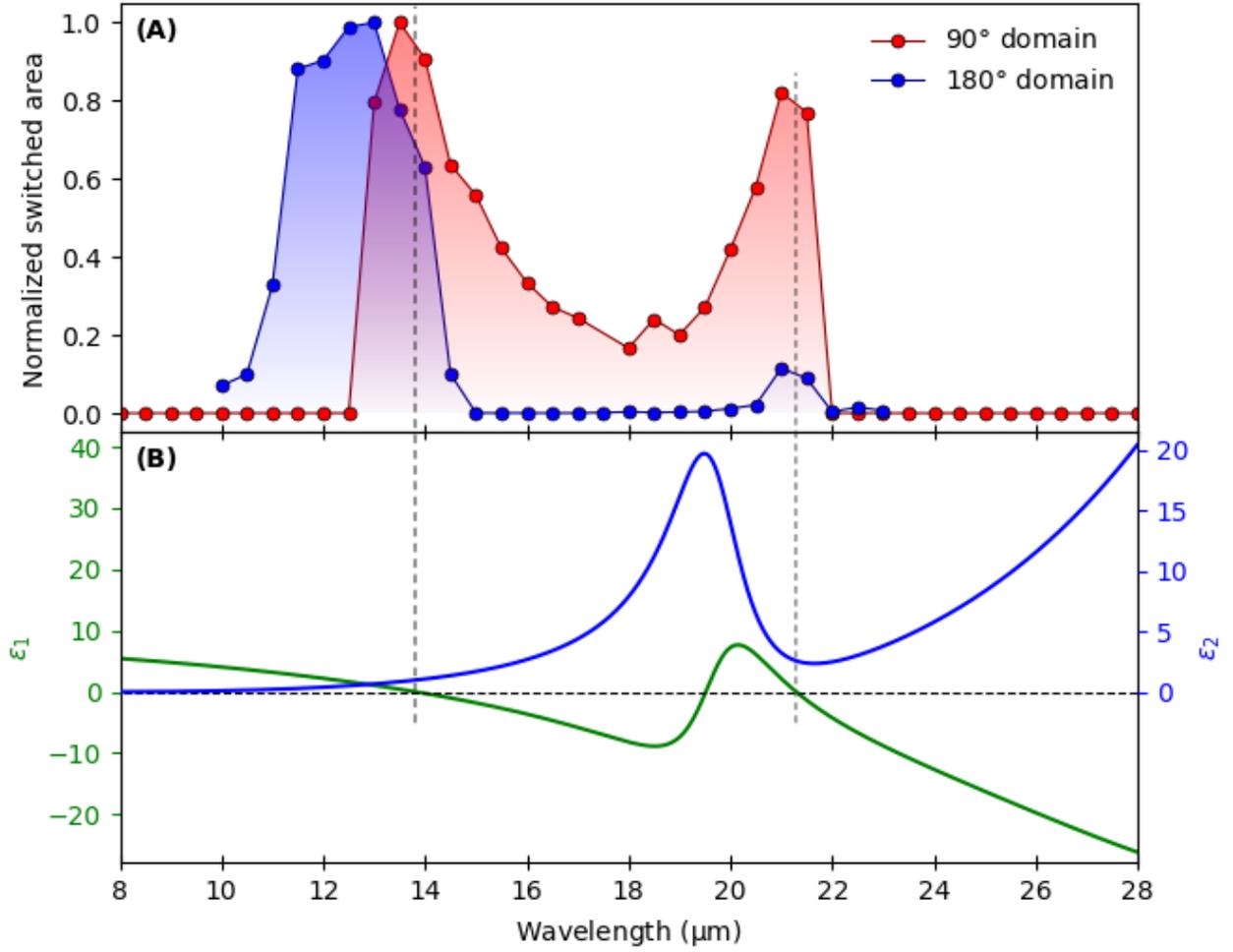

**Fig 2: Importance of satisfying the ENZ condition for switching.**
**(A)** The spectral dependence of the switching efficiency, evaluated as the normalized area of the switched 90° and 180° domains (red and blue points respectively). **(B)** The spectral dependence of the real and imaginary parts of the permittivity $\varepsilon = \varepsilon_1 + i\varepsilon_2$ of $BaTiO_3$, taken from Ref. [17]. The switching is maximized in strength at the ENZ condition, i.e., when both $\varepsilon_1$ and $\varepsilon_2$ approach zero.

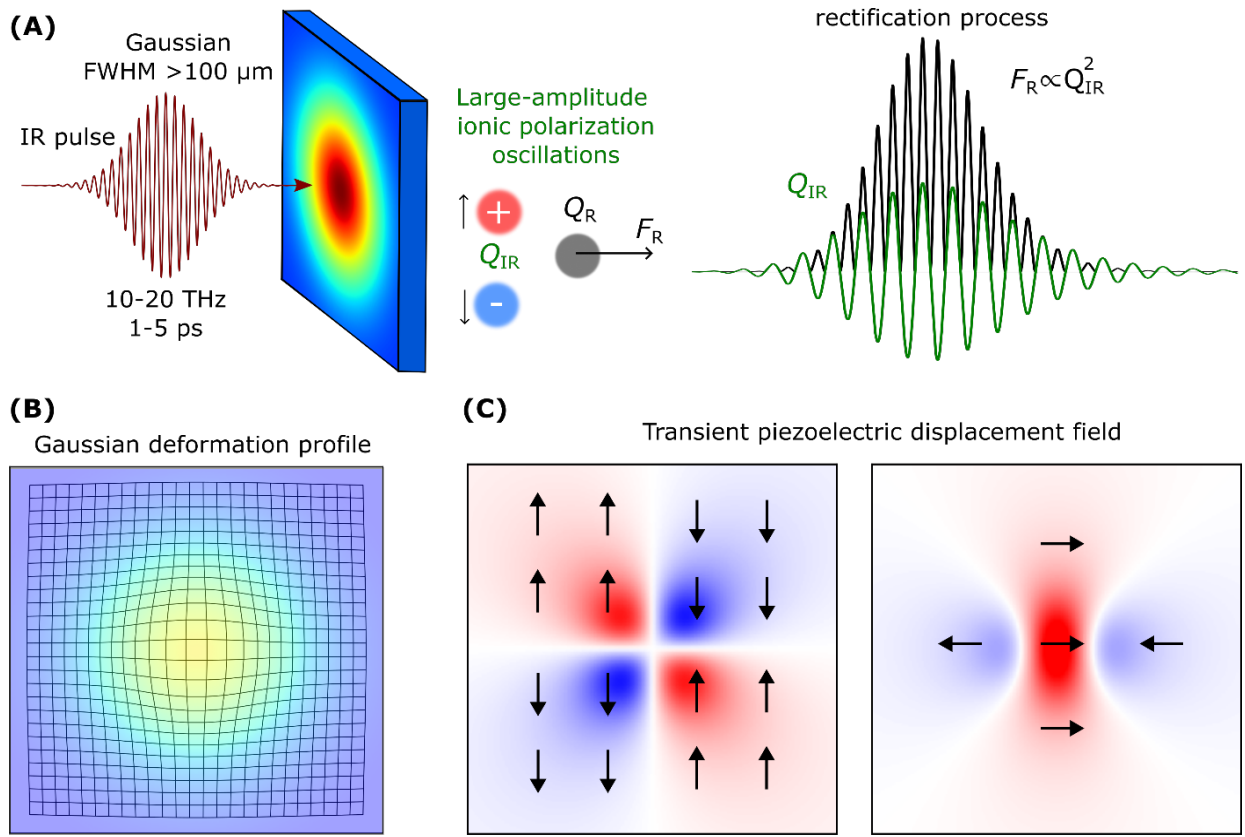

**Fig. 3: Schematic explanation of the switching process.**
**(A)** An ultrafast IR pulse, tuned in frequency to meet the ENZ condition, induces large-amplitude ionic polarization oscillations along the coordinate $Q_{IR}$ that in turn non-linearly drives a rectified force ($F_R$) acting on the atoms along the coordinate $Q_R$. In this case, we show the rectified force that would emerge if non-linear phononics would be the cause. **(B)** The rectified force on the atoms leads to a deformation profile that follows the Gaussian intensity profile of the impinging IR pulse. **(C)** The resulting calculated piezoelectric displacement field that gives rise to 90° and 180° switched domains (left and right panels respectively), assuming that the ferroelectric polarization initially points to the right.

# Methods:

## Materials:

The barium titanate crystals studied in our experiments were procured from MSE Supplies. The crystals, of size 5×5×0.5 mm$^3$, were grown with the Czochralski process, and both surfaces were polished. We selected a (100) crystal orientation so that the spontaneous ferroelectric polarization lays along the crystal surface.

## Pumping of BaTiO$_3$:

To explore the possibility of switching ferroelectric polarization at the ENZ conditions, we used pump laser pulses delivered by the free-electron laser facility FELIX in Nijmegen, The Netherlands. The temporal structure of the IR pulses is shown in Supplementary Data Fig. S1. The free-electron laser supplies transform-limited IR pulses at a repetition rate of 25 MHz. These pulses come within 8-µs-long bursts ("macropulses") at a rate of 10 Hz. The central wavelength of the pump pulses, with wavelength-dependent durations in the range 1−3 ps, was varied in the spectral range of 7−28 µm with their bandwidth experimentally tunable in the range of 0.5−2.0% (typically <1% of the bandwidth used in our experiments). The laser beam was focused to a spot with a diameter of about 200 µm onto the surface of the BaTiO$_3$ crystal.

## Observation of polarization domains by microscopy:

In order to observe the ferroelectric domains and their dynamics, we have used (i) a polarization microscope and (ii) a second harmonic generation (SHG) microscope. The polarization microscope uses the birefringence in the material to distinguish different domains[16]. Linearly-polarized light, of wavelength $\lambda$ = 520 nm as delivered by a CW diode laser (Integrated Optics), illuminates the sample at normal incidence. The transmitted light is collected by an objective lens and directed through an analyser (i.e., polarizer) in order to select the transmitted component polarized along a specific axis. After such filtering, the light is spatially-resolved by the matrix of a charge-coupled device (CCD) camera. The CCD is suitably triggered to record an image before and after arrival of the pump pulse. The contrast of this microscope stems from the fact that the refractive index along the c-axis (the crystal axis parallel to the ferroelectric polarization) differs from that associated with the other crystal axes (a-axes). Thus, we are able to distinguish 90° domains using polarization microscopy but we are unable to distinguish 180° domains.

To distinguish 180° domains, we use SHG microscopy. We illuminate the sample with a train of ≈150-fs-long laser pulses, of wavelength $\lambda$ = 1040 nm, at a repetition rate of 100 MHz (Menlo Orange HP). Since our probe pulse energies (up to 90 nJ) are capable of burning the sample, we use an optical chopper to modulate the illumination and thus reduce the optical power. The transmitted 1040-nm and SHG-generated 520-nm light is collected by an objective lens, and the former is extinguished by an appropriate short-pass filter. The SHG is then recorded by a CCD. The SHG signal produced in BaTiO$_3$ depends on the non-linear tensor components of the $\chi^{(2)}$ tensor. The signal generated from different domains differ since different components of the $\chi^{(2)}$ tensor are used in the generation of the second harmonic[31]. This makes it quite straightforward to distinguish 90° domains. Observation of 180° domains however is less trivial and we have to realize that the created domains by FELIX only appear close to the surface and are not going through the entire bulk of the crystal. The SHG signal of 180° domain is equal in intensity but with a 180° phase difference. Due to this phase difference and the fact that there is no phase matching in BaTiO$_3$, the SHG signal of two successive 180° domains can constructively interfere and appear brighter compared to a single bulk domain[32].

## Model of the switching process:

Here we elaborate our model for the switching process, which was schematically shown in Fig. 3. We assume the free energy $F$ of BaTiO$_3$ can be expanded in the following form[33]:

$$F = F_0 + \mu\left(u_{ik} - \frac{1}{3}\delta_{ik}u_{ll}\right)^2 + \frac{1}{2}Ku_{ll}^2 + \frac{1}{2}K_i Q_i^2 + \alpha_{ijkl}\, u_{ij}Q_k Q_l \tag{1}$$

where the first three terms are of pure elastic origin, assumed to be isotropic. Following the notation of Ref. 33, $\mu$ is the modulus of rigidity, $u_{ik}$ the strain tensor, $K$ the modulus of compression and $Q$ the phonon coordinate. The fourth term, quadratic in $Q_i$, is the energy of the laser-excited phonon modes. The last term is a coupling between the phonon coordinate and the strain[34], and is of particular importance for explaining our results. We assume that the laser excites only a single phonon mode ($Q_k = Q_l = Q_{IR}$). In this case, the excitation can only lead to volumetric changes due to the symmetry restrictions of the 4mm point group. Furthermore, to make calculations analytically solvable, we assume that the coupling term between the strain and the phonon coordinate is in the form $\alpha_{ijkk} = \alpha\delta_{ij}$, i.e., the strain coupling is equal in every direction. In general, this term could of course vary, but we find that this assumption of isotropy is sufficient to explain the experimentally-observed switching of ferroelectric polarization. We note that no ab-initio calculations were performed during this study to obtain the real values of these coefficients.

We will now focus on the effects of the last term in Eq. (1), and elaborate how it can create a Gaussian deformation profile. With the stress tensor being given by $\sigma_{ik} = \frac{\partial F}{\partial u_{ik}}$, the force acting on a volume element is given by the divergence of the stress tensor $F_i = \frac{\partial \sigma_{ik}}{\partial x_k}$. The force on a volume element therefore becomes $\vec{F}(r) = \alpha\nabla Q_{IR}^2$. Since the phonon coordinate scales proportionally with the laser's electric field, $Q_{IR}^2$ scales proportionally with the intensity ($I$) of the laser pulse, and thus $\vec{F}(r) \propto \alpha\nabla I(r)$.

With the above information, we can calculate the pulse-induced deformation/displacement ($\vec{u}$) in the material using the following equation (*33*):

$$\nabla\cdot\nabla\cdot\vec{u} - \frac{1-2\sigma}{2(1-\sigma)}\nabla\times\nabla\times\vec{u} = \frac{(1+\sigma)(1-2\sigma)}{E(1-\sigma)}\vec{F} \tag{2}$$

where $E$ is the Young's modulus. The applied laser pulse has a Gaussian spatial profile $I(r) = A\cdot e^{-r^2/2a^2}$. Thus, the displacement will be purely radial and the term $\nabla\times\vec{u}$ can be dropped. Solving the above equation gives

$$\vec{u}(r) = \frac{1}{r}\beta A a^2\left(1 - e^{-\frac{r^2}{2a^2}}\right) \tag{3}$$

with:

$$\beta = \frac{(1+\sigma)(1-2\sigma)}{E(1-\sigma)}\alpha \tag{4}$$

This leads to the strain profile:

$$u_{xx} = \frac{\beta A a^2}{x^2 + y^2}\left[\left(1 - e^{-\frac{x^2+y^2}{2a^2}}\right)\left(1 - \frac{2x^2}{x^2+y^2}\right) + \frac{x^2}{a^2}e^{-\frac{x^2+y^2}{2a^2}}\right] \tag{5}$$

$$u_{yy} = \frac{\beta A a^2}{x^2 + y^2}\left[\left(1 - e^{-\frac{x^2+y^2}{2a^2}}\right)\left(1 - \frac{2y^2}{x^2+y^2}\right) + \frac{y^2}{a^2}e^{-\frac{x^2+y^2}{2a^2}}\right] \tag{6}$$

$$u_{xy} = -\frac{\beta A a^2 xy}{x^2 + y^2}\left[\frac{2}{x^2 + y^2}\left(1 - e^{-\frac{x^2+y^2}{2a^2}}\right) - \frac{1}{a^2}e^{-\frac{x^2+y^2}{2a^2}}\right] \quad (7)$$

In ferroelectrics this strain may couple to the polarization via the piezoelectric effect. In barium titanate in particular, strain ($u$) can induce an electric displacement field ($D$) via the piezoelectric effect as following[35]:

$$\begin{pmatrix}D_1\\D_2\\D_3\end{pmatrix} = \begin{pmatrix}0 & 0 & 0 & 0 & e_{15} & 0\\0 & 0 & 0 & e_{15} & 0 & 0\\e_{31} & e_{31} & e_{33} & 0 & 0 & 0\end{pmatrix}\begin{pmatrix}u_1\\u_2\\u_3\\u_4\\u_5\\u_6\end{pmatrix} \quad (8)$$

With $e_{15} = 68.4 \frac{C}{m^2}, e_{31} = -0.7 \frac{C}{m^2}, e_{33} = 6.7 \frac{C}{m^2}$. Using these values, we have calculated the polarization created by the Gaussian strain profile. The symmetry of this polarization profile is shown in the Fig. 3(C) and it matches that of the switched domains.

**Data availability**

The data supporting the findings of this study are available within the article and its supplementary file. Additional data can be obtained from the authors upon a reasonable request.

## Supplementary information

### Heating effects

The absorption of the laser pulse by BaTiO$_3$ leads to the generation of heating. This heat has an influence on the refractive index $n$ and birefringence ($\Delta n$) of barium titanate[1], and thus its effects are visible through the polarization microscope. The Gaussian intensity profile of the laser pulse leads to a Gaussian profile of heating in the sample, and thus a Gaussian spatial profile of retardation since the retardation is given by $\delta = \Delta n(T) \cdot d$ where $d$ is the thickness of the crystal. Using data from Ref.1, we estimate the total retardation of our 0.5 mm-thick sample is in the order of 50 waves at room temperature. Heating the sample can significantly decrease this retardation.

We now consider how this heat-dependent retardation profile will manifest in the polarization microscope. To understand this, we note that the polarization microscope can only distinguish between relative retardations, rather than the absolute retardation (e.g., a retardation of 46.5 waves will appear the same as 47.5 waves). Let us consider the images in Fig. S2(A). Light and dark contrast corresponds to varying amounts of retardation, with the ringed structure stemming from the radial symmetry of the Gaussian heat profile. Each ring is equivalent to one full wave of retardation difference, and so the number of rings gives an estimate of the amount of heat absorbed from the laser pulse. More rings therefore corresponds to a stronger thermal profile.

Upon varying the pump wavelength, we observe that the number of rings vary. We therefore counted the number of rings generated using excitation wavelengths between 8 and 14 μm (Fig. S2(B)) and discovered that the switching observed at λ = 14 μm is actually not correlated with a large temperature change of the sample. Moreover, the incident pulse energy does not vary significantly across this spectral range. The slightly-rectangular shape of the rings and the lobes, mainly visible at lower wavelengths, can be explained by including the effect of strain-induced birefringence induced by thermal expansion.

### Lifetime of transient domains

Under certain conditions, we frequently observed the rapid decay of laser-induced switched domains of ferroelectric polarization. To measure the lifetime of these transient domains, we repeated the microscopy measurements with temporal-resolution provided by synchronizing the CCD camera with the macropulses of the free-electron laser. The exposure time of the CCD – 27 μs – defines the time resolution. While the appearance of switched domains is instantaneous on this timescale, we observe that a large number of domains shrink and vanish across a timescale of milliseconds (the exact time-scale depends on the size of the domain). The shrinking and disappearance of newly-formed switched domains probably stems from neighboring dipole-dipole interactions. In addition, the created domains do not extend through the whole thickness of the crystal. When a domain achieves a critical size, however, it may become stable. In some situations, with the appropriate wavelength and with sufficiently strong pulse intensity, we observed that the switched domains did not shrink and thus were stable. The position of the laser spot seemed to play a potentially significant role on the domain formation process. At certain places on the sample, e.g., near defects or at the edges of the sample, it seemed easier to create domains.

In Figs. S3-S4, we present a series of images taken using polarization and SHG microscopy respectively. These two figures respectively show the dynamics of 90°- and 180°-domains. The images shown were taken at the time-delays *t* indicated relative to the arrival of the IR macropulse. The images have been background-corrected, i.e., an image taken before the arrival of the IR pump has been subtracted, to emphasize the laser-induced features. Grey contrast implies there is no intensity difference between the two images.

**Image of coexisting 90° and 180° switched domains**

Figure S5 shows a representative image taken in which a single pumping macropulse has switched both 90° and 180° domains simultaneously (circled orange and green respectively). This image was obtained using SHG microscopy. This image confirms that the piezoelectric displacement fields that are induced by the IR pump pulse, as shown in Fig. 3(C), coexist.

**Dependence of switching on pulse energy**

We have performed additional measurements assessing how the size of the 180° and 90° switched domains scale with the energy of the infrared laser pulse. The results are shown as the blue points in Fig. S6. We have fitted the measured data points with an exponential function of the form $a \cdot e^{-b/U_{pulse}}$. This dependence is in accordance with our model for the switching mechanism, since the strain induced by a laser pulse scales linearly with the pulse energy. In turn, the electric displacement field created by the piezoelectric effect is linearly proportional to the strain. Hence, the laser-induced electric displacement field is linearly proportional to the pulse energy. Experiments studying the field-driven formation of domains, and the subsequent domain-wall motion, have shown that the switching rate of 180° domains increases exponentially with the exponent $-E_a/E$ where $E_a$ is the 'activation field' and $E$ is the electric field[2]. Our model of the switching consequently implies that the domain size scales exponentially with the form $a \cdot e^{-b/U_{pulse}}$.

While the exponential function clearly fits well the size of the 180° domains, the fit is less successful in describing the size of the 90° domains. This difference can be attributed to the fact that the switching takes place by the inhomogeneous nucleation of many small domains[2]. For the 180° domains, many (uncountable) domains can be created by a laser pulse, while more pulse energy is required to switch 90° domains and it was only possible to create up to 3 switched domains with a single macropulse. This limited nucleation makes it possible to only observe the 'linear part' of the exponential function, in the case of the 90° domains.

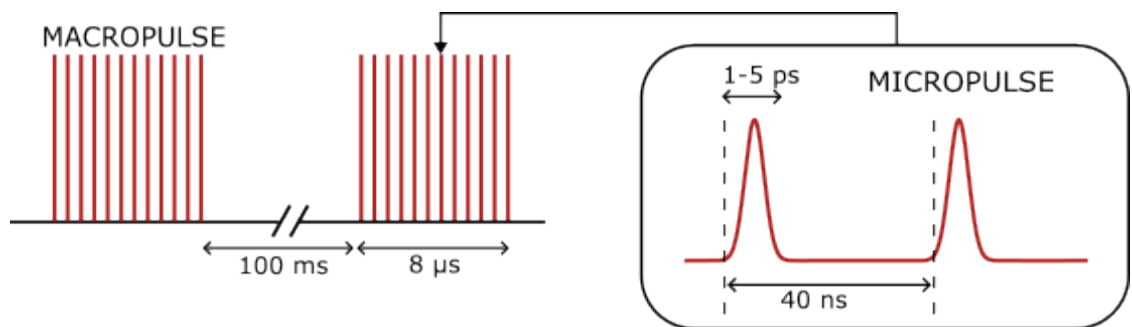

**Fig. S1: Temporal structure of the pump pulses.**

The free-electron laser supplies transform-limited infrared micropulses at a repetition rate of 25 MHz. These pulses come within 8-µs-long bursts ("macropulses") at a rate of 10 Hz.

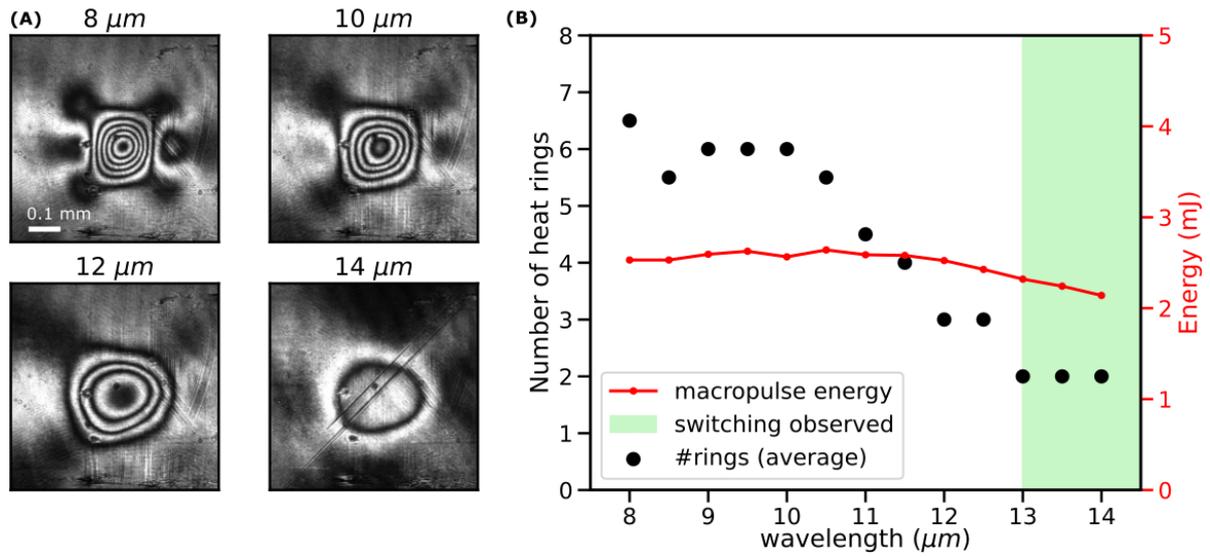

**Fig. S2: Heating of sample by the laser pulse**

**(A)** Some images of the heating effects at different wavelengths just after application of the laser pulse. Note that switching (the diagonal stripes) is only visible at the image of 14 μm. **(B)** Wavelength dependence of the heating effects and domain switching.

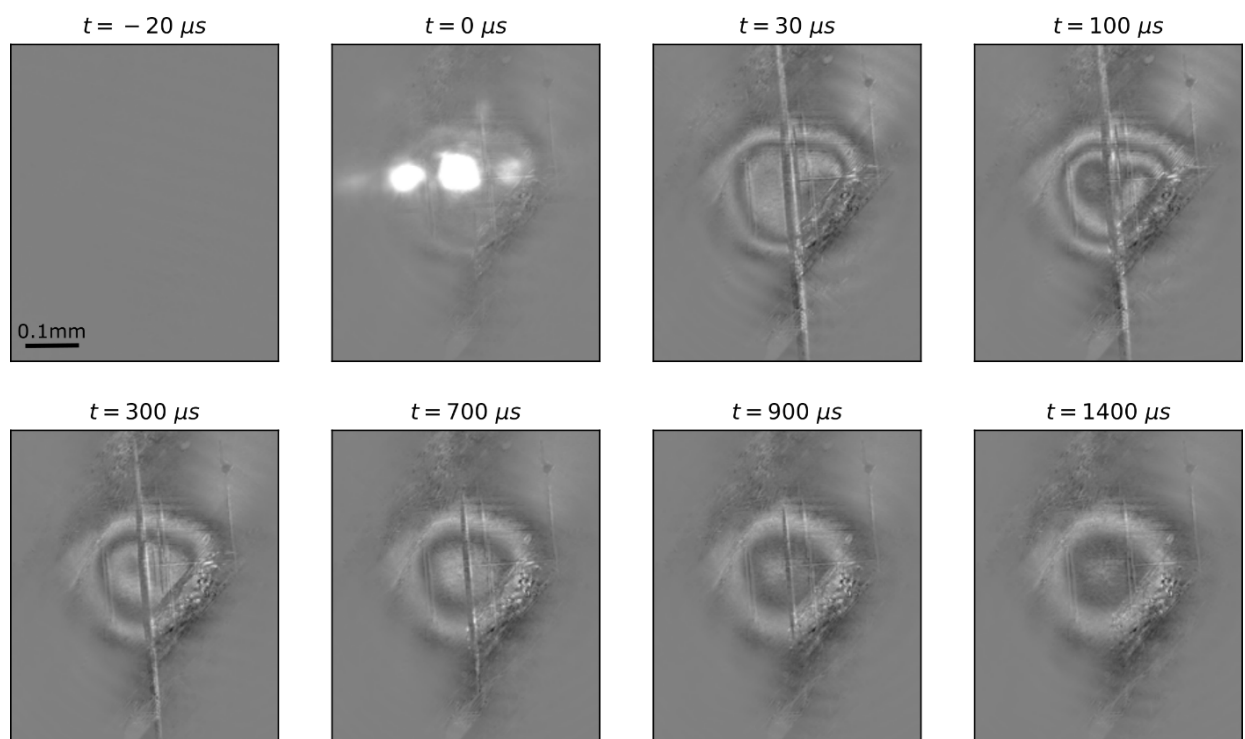

**Fig. S3: Time dependence for 90° switched domains**

Time evolution of the switching process observed with polarization microscopy. At t = 0 µs, we observe a bright spot which is an artefact coming from higher harmonics created by the IR laser pulse in the system. After the laser pulse we see a stripe, which contrast corresponds to a 90° domain that has been created. We also see rings coming from heating effects.

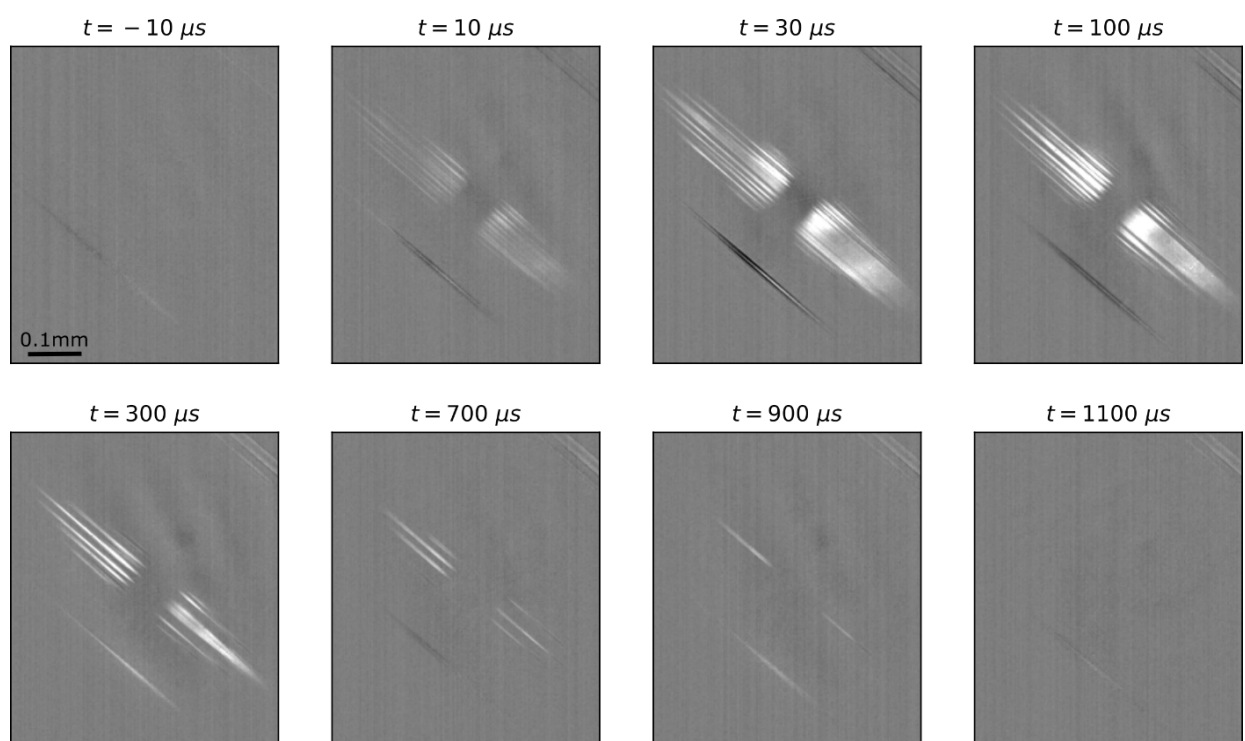

**Fig. S4: Time dependence for 180° switched domains**

Time evolution of the switching process observed with second harmonic generation microscopy. The bright stripes are 180° domains created by the laser pulse.

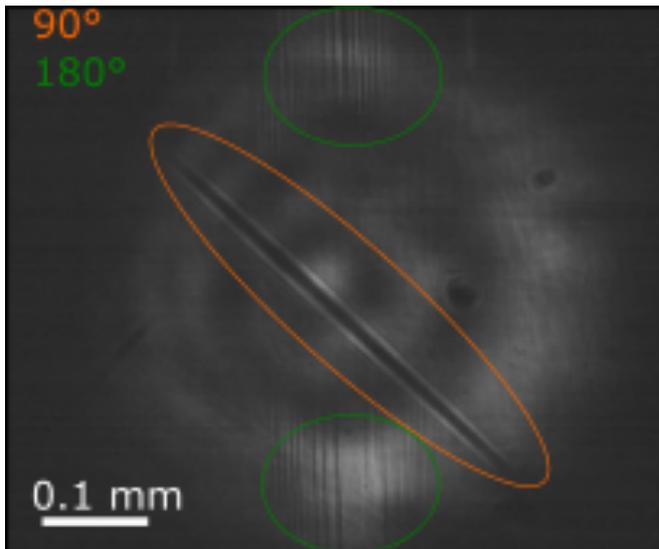

**Fig. S5: Simultaneous switching of 90° and 180° domains**

Picture of switched domains, taken 30 µs after pumping BaTiO3 with a macropulse of wavelength 21.5 µm. The picture was taken using SHG microscopy. The 90° and 180° domains are circled orange and green respectively.

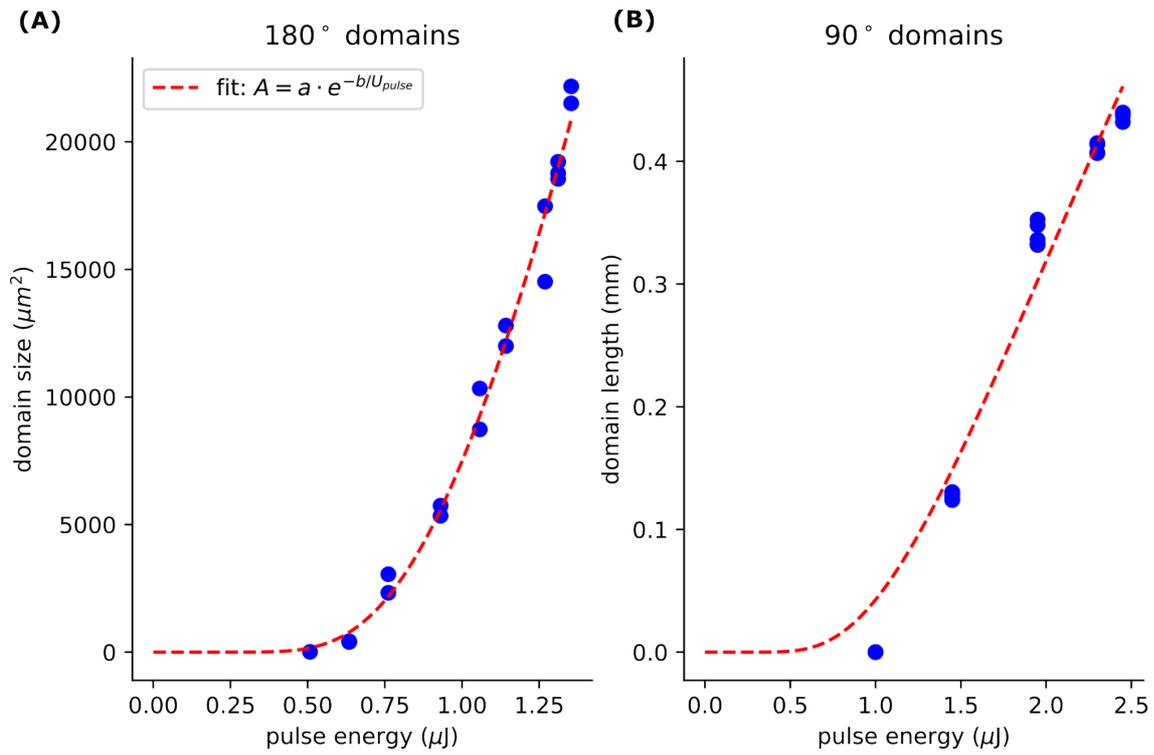

**Fig. S6: Dependence of switched domains on the pulse energy.**

**(A)** Using SHG microscopy, we imaged the 180° domains switched by a pump pulse of wavelength 12 μm with varying energy. Since the switched 180° domains consisted of very closely packed lines, we measured the total spatial extent of these domains. **(B)** Using polarizing microscopy, we imaged the 90° domains switched by pump pulses of varying energy. This time the pump pulse has a wavelength of 14 μm. Since there were relatively less domain lines, we measure the total length of these domains. In both panels, the blue points correspond to the measurements, and the red dashed line corresponds to an exponential fit as described in the accompanying text.